\documentclass{ws-ijmpcs}
\usepackage[super,compress]{cite}
\usepackage[british]{babel}
\usepackage{xspace}

\begin{document}

\markboth{Sara Pucillo for the ALICE Collaboration}{Recent results on strangeness enhancement in small collision systems with ALICE}

%
\catchline{}{}{}{}{}
%

\title{Recent results on strangeness enhancement in small collision systems with ALICE}

\author{Sara Pucillo \footnote{sara.pucillo@cern.ch} for the ALICE Collaboration}

\address{Dipartimento di Fisica, Università degli Studi di Torino and Sezione INFN Torino, \\Via Pietro Giuria 1, Torino, Italy\\
sara.pucillo@unito.it}

\maketitle


\begin{abstract}
\noindent Quantum Chromodynamics (QCD) predicts that, at sufficiently high temperature and energy density, nuclear matter undergoes a phase transition from confined hadrons to a deconfined state of quarks and gluons known as the quark-gluon plasma (QGP). One of the historically proposed signatures of QGP formation is strangeness enhancement (SE), characterized by an increased production of strange hadrons in heavy-ion collisions relative to proton--proton (pp) interactions. At the LHC, the ALICE experiment has measured a continuous increase in the strange-to-non-strange hadron yield ratios as a function of midrapidity charged-particle multiplicity, not only in large systems like Pb--Pb but also in small systems such as pp and p--Pb. The origin of SE in small systems is still under debate, motivating further experimental investigations. This article presents recent ALICE analyses that offer complementary insights into the phenomenon. These include (i) multi-differential studies using event-shape observables such as transverse spherocity and the concept of effective energy, and (ii) the first measurement of multiplicity distributions of strange and multi-strange hadrons, P($\textit{n}_{S}$), in pp collisions.
\end{abstract}

\section{Introduction}

\noindent The measurement of strange hadron production is fundamental to understanding the hadronization mechanisms and the properties of the strongly interacting medium created in high-energy collisions. An enhanced production of strange particles, particularly multi-strange baryons, was initially proposed as a signature of QGP formation~\cite{Rafelski:1982}. This effect was first observed at the CERN SPS~\cite{Andersen:1999} and subsequently confirmed at RHIC~\cite{starSE} and the LHC~\cite{PbPb2}. ALICE has extended these studies across multiple collision systems and energies~\cite{nature,pp7,pp13}, revealing a common trend: the relative production of strange hadrons increases with event multiplicity. This enhancement persists even in high-multiplicity pp collisions, which do not form extended QGP-like media according to conventional expectations. The strangeness content of the hadron correlates with the magnitude of the enhancement, with the Ω baryon showing the strongest effect.
\noindent While this trend is well-established, current QCD-based models struggle to reproduce the observed patterns quantitatively. Theoretical frameworks such as a statistical hadronisation description using the canonical suppression approach~\cite{canonical}, rope hadronisation models including colour reconnection effects~\cite{ropecit}, and two-component (core-corona) models~\cite{corecit} offer partial explanations. However, a full understanding requires further exploration of the event properties that influence strangeness production. In this context, ALICE has explored the role of event topology~\cite{spherocity} and effective energy~\cite{effective} and has pioneered the measurement of strange hadron multiplicity distributions on an event-by-event basis~\cite{mythesis}.

\section{Experimental apparatus}

\noindent The ALICE detector is optimized for heavy-ion physics and provides excellent particle tracking and identification over a broad momentum range. A complete description of the detector performance during Runs 1 and 2 is available in Refs.~\citen{ALICEjinst, ALICEperf}; here, the main detectors relevant for the current analyses are briefly described. The Inner Tracking System (ITS~\cite{ALICEits}) and Time Projection Chamber (TPC~\cite{ALICEtpc}), both operating within a solenoidal magnetic field of 0.5 T, offer tracking and particle identification (PID) at midrapidity ($|\eta| < 0.9$). The ITS is composed of six layers of silicon detectors (pixel, drift and strip), providing precise vertexing and low-$p_{\rm T}$ tracking. The TPC records up to 159 tracking points per particle and delivers PID via specific energy loss ($\mathrm{d}E/\mathrm{d}x$). The Time-Of-Flight (TOF~\cite{ALICEtof}) detector complements PID performance in the intermediate $p_{\rm T}$ range. Strangeness is reconstructed via weak decay topologies within the central rapidity region ($|y| < 0.5$). Strange hadrons such as K$^{0}_{S}$, $\Lambda$, $\overline{\Lambda}$, $\Xi^{-}$, $\overline{\Xi}^{+}$, $\Omega$ and $\overline{\Omega}^{+}$ are identified using geometric selection criteria and invariant mass analysis. Additional detectors include the V0~\cite{ALICEv0} scintillators (V0A and V0C), covering $2.8 < \eta < 5.1$ and $-3.7 < \eta < -1.7$, respectively, used for triggering and multiplicity classification. The Zero Degree Calorimeters (ZDC) measure neutral and charged baryons at very forward rapidity and are key to estimating the effective energy.

\section{Multi-differential analyses of strangeness production in small collision systems}

\noindent In order to explore the origins of SE in small collision systems more thoroughly, the ALICE Collaboration investigates whether factors beyond the average charged-particle multiplicity, $\langle$d$N_{ch}$/d$\eta \rangle$, can account for the observed behaviors. This method involves categorizing high-multiplicity (HM) events based on event topology using transverse spherocity~\cite{spherocity} (see Section ~\ref{capspherocity}), as well as isolating global properties from local effects by introducing the concept of effective energy~\cite{effective} (see Section ~\ref{capeffective}) in pp collisions at $\sqrt{s} = 13$~\rm{TeV}.

\subsection{Event topology with transverse spherocity} \label{capspherocity}

\noindent Event topology can offer information on the underlying particle production mechanisms~\cite{azimuthal}. One useful observable to characterize this topology is the unweighted (transverse momentum-independent) transverse spherocity, $S_{0}^{p_{\rm T} = 1}$, defined as:

\begin{equation}
 S_{0}^{p_{\rm T} = 1} = \frac{\pi^{2}}{4} \min_{\hat{n}} \left( \frac{\sum_{i} |\hat{p_{\rm T}}_{i} \times \hat{n}|}{N_{\rm trks}} \right)^{2}
 \label{eq:spherocitydef}
\end{equation}

\begin{figure}[htbp]
    \centering
    \includegraphics[height=0.45\linewidth]{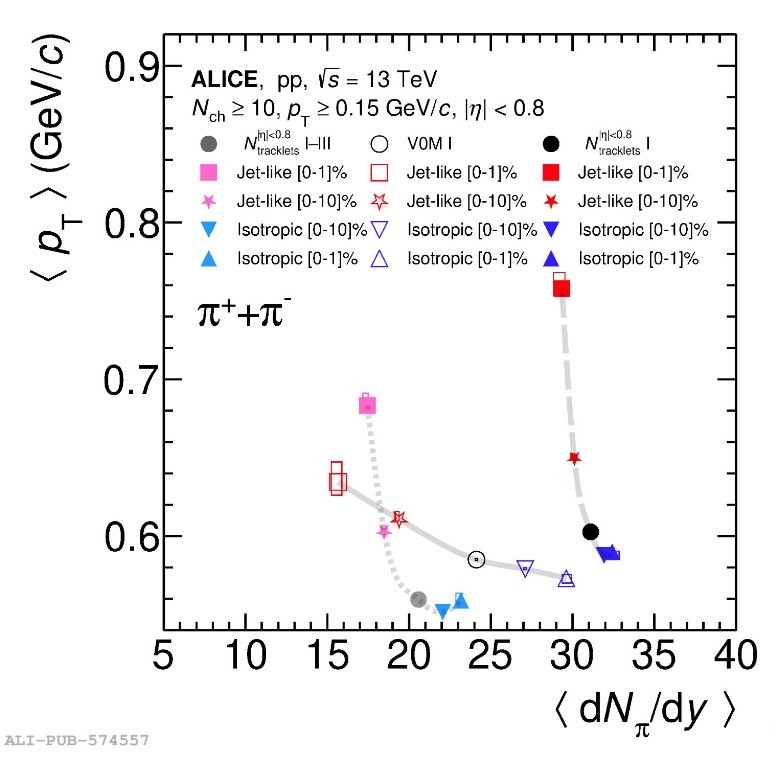} 
    \hspace*{0.2cm}
    \includegraphics[height=0.45\linewidth]{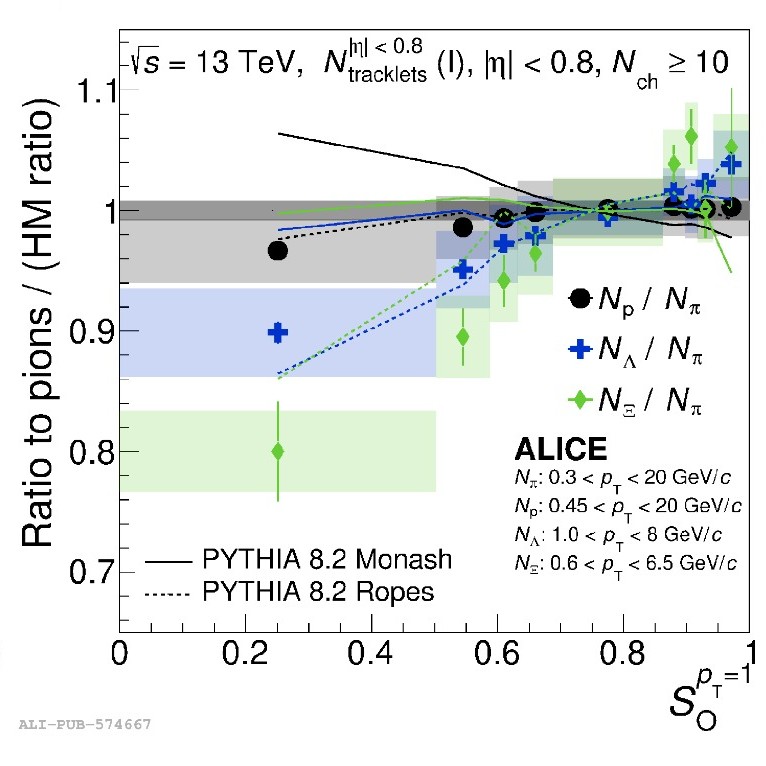}
    \caption{\textit{Left.} Correlation between average transverse momentum $\langle p_{\rm T} \rangle$ and the average pion yield $\langle$d$N_{\pi}$/d$y \rangle$ as a function of $S_{0}^{p_{\rm T} = 1}$, in the 0-10\% and 0-1\% $N_{\rm tracklets}^{|\eta|<0.8}$ and V0M multiplicity classes~\cite{spherocity}. \textit{Right.} Integrated yields of p, $\Lambda$ and $\Xi$ as a function of $S_{0}^{p_{\rm T} = 1}$ for the $N_{\rm tracklets}^{|\eta|<0.8}$ 0-10\%, relative to the corresponding $\pi^{-}$+$\pi^{+}$ yields. Model predictions have been reported for PYTHIA 8.2 Monash and Ropes as continuous and dotted lines respectively.}
    \label{fig:Spherocity}
\end{figure}

\noindent In this expression, {\ensuremath{p_{\rm T}}\xspace} denotes the transverse momentum unit vector of each detected particle in the azimuthal plane, $N_{\rm trks}$ represents the number of reconstructed primary charged tracks in the event and $\hat{n}$ is the unit vector that minimizes $S_{0}^{p_{\rm T} = 1}$. The transverse spherocity, $S_{0}^{p_{\rm T} = 1}$, ranges from 0 (jet-like, back-to-back events) to 1 (isotropic distributions), and is sensitive to the relative contribution of hard and soft processes. By studying $S_{0}^{p_{\rm T} = 1}$ in high-multiplicity pp events at $\sqrt{s} = 13$~\rm{TeV}, ALICE investigates whether SE is driven by specific topological classes~\cite{spherocity}. In pp collisions, the presence of multiple independent parton–parton interactions can naturally lead to a more isotropic azimuthal distribution as the number of scatterings increases. As such, before examining the production of strange particles as a function of $S_{0}^{p_{\rm T} = 1}$, it is crucial to consider potential biases that may result from this trivial isotropization in events with high multiplicity. This effect is investigated in Fig.~\ref{fig:Spherocity} (left panel), which displays the average transverse momentum $\langle p_{\rm T} \rangle$ and the average pion yield $\langle$d$N_{\pi}$/d$y \rangle$, across various spherocity and multiplicity intervals~\cite{spherocity}. When using central multiplicity estimators, denoted as $N_{\rm tracklets}^{|\eta|<0.8}$ (the top 10\% (I–III) and 1\% (I) of events with the highest number of short track segments-tracklets- measured in the ITS within $|\eta|<0.8$), the average {\ensuremath{p_{\rm T}}\xspace} varies significantly with $S_{0}^{p_{\rm T} = 1}$, indicating a change in the event hardness. In contrast, when using forward multiplicity (V0M), the average {\ensuremath{p_{\rm T}}\xspace} remains nearly constant with $S_{0}^{p_{\rm T} = 1}$. This confirms that combining central multiplicity and spherocity can better disentangle contributions from hard and soft processes. The right panel of Fig.~\ref{fig:Spherocity} shows the integrated, self-normalized ratios of strange hadrons to pions—specifically for p, $\Lambda$ and $\Xi$-as a function of $S_{0}^{p_{\rm T} = 1}$, within the top 0–1\% of events selected using $N_{\rm tracklets}^{|\eta|<0.8}$. These results show suppression in jet-like ($S_{0}^{p_{\rm T} = 1} \rightarrow 0$) and mild enhancement in isotropic ($S_{0}^{p_{\rm T} = 1} \rightarrow 1$) events. This pattern scales with strangeness content and supports a multi-source scenario for particle production. When compared to theoretical predictions, while PYTHIA 8.2 with rope hadronization~\cite{PythiaRopes} reproduces the general trend, it fails to capture the strangeness hierarchy; the Monash tune~\cite{Pythia} underestimates all ratios.

\subsection{Effective energy and initial state effects} \label{capeffective}

\begin{figure}
    \begin{center}
    \includegraphics[width=0.99 \textwidth]{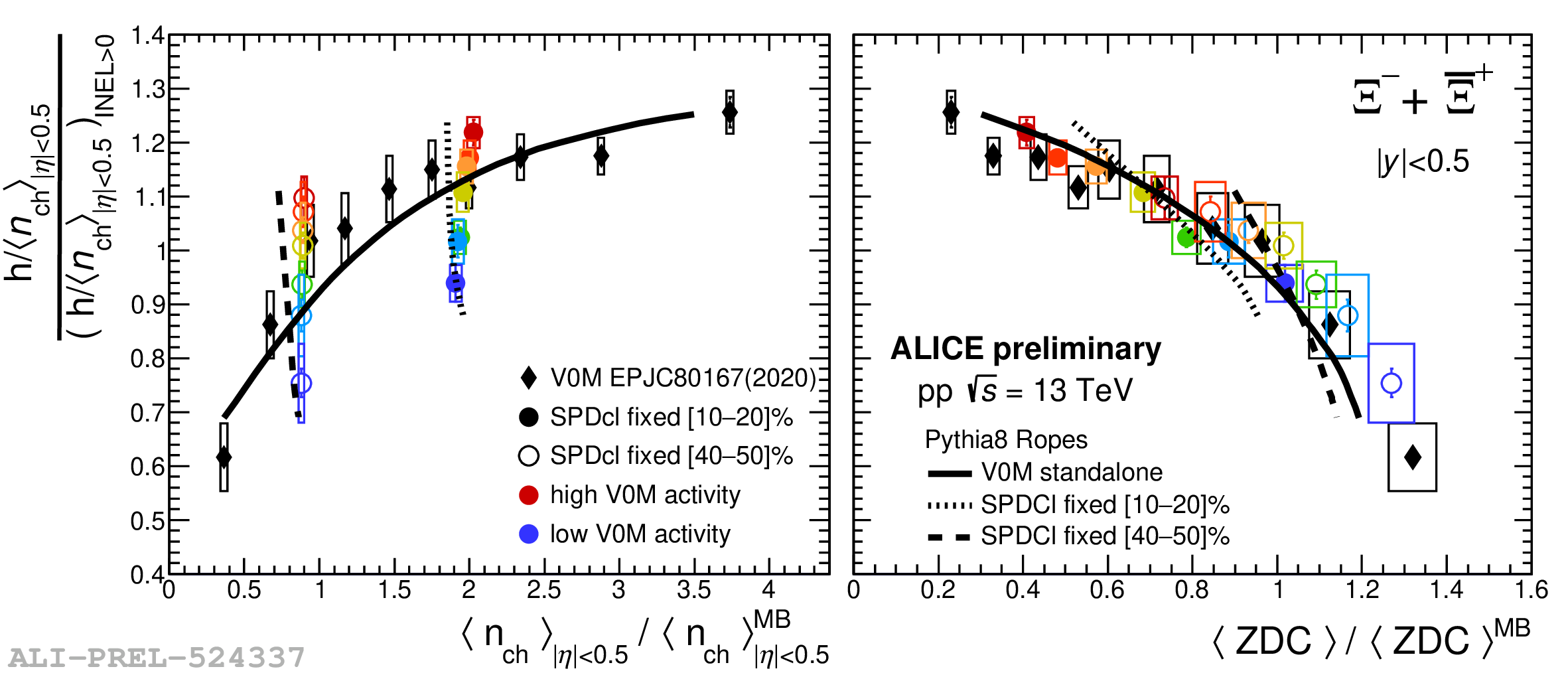}
    \end{center}
    \caption{Self-normalised $\Xi$ yields per charged particle as a function of the particle multiplicity produced at midrapidity (left) and of the very forward energy detected by the ZDC (right). The V0 standalone selection is shown in black diamonds and coloured circles show the “fixed multiplicity” events selected using the information from the SPD and V0 detectors. The Pythia8 Ropes predictions are shown with full and dashed lines, respectively.}
    \label{fig:effectiveenergy}
\end{figure}

\noindent In pp collisions, charged-particle multiplicity characterizes the final state, which is also closely linked to the initial energy available for particle production. To separate initial- from final-state effects, ALICE introduces the concept of effective energy—estimated by subtracting forward energy (measured by ZDCs) from the nominal $\sqrt{s}$. Midrapidity multiplicity shows an anticorrelation with ZDC energy~\cite{effectiveenergy}, indicating a connection between forward baryon transport and central particle production. This study~\cite{effective} aims at disentangling the respective contributions of midrapidity multiplicity and initial effective energy to strange particle production. Two event classes are analyzed: standalone (based on V0 signal amplitude) and fixed multiplicity (selected with both SPD-the two innermost layer of the ITS- and V0), allowing selection of events with similar midrapidity multiplicities but varying forward energy. Strange hadron production is then measured in each class and yield ratios, normalized to charged-particle multiplicity, are used as a proxy for the strange-to-non-strange particle ratio. Figure~\ref{fig:effectiveenergy} presents the self-normalized $\Xi$ yields per charged particle in standalone and fixed multiplicity events. The left panel shows dependence on midrapidity charged-particle production, while the right panel displays dependence on forward energy. As previously observed~\cite{nature}, the standalone selection confirms an increase in strange hadron yields with multiplicity. Additionally, a decrease in $\Xi$ production is seen with increasing forward energy. In fixed-multiplicity classes, a clear enhancement in $\Xi$ yields per charged particle is observed as the forward energy decreases, suggesting that effective energy influences strangeness production. This behavior appears to follow a universal trend across event classes. Comparisons with theoretical models show that Pythia 8 with the Ropes mechanism~\cite{PythiaRopes} qualitatively reproduces the observed rise in $\Xi$ yields with decreasing forward energy.

\section{(multi-)strange particle multiplicity distribution} \label{cappns}

\begin{figure}
    \centering
    \includegraphics[width=0.45 \textwidth, height=0.45\linewidth]{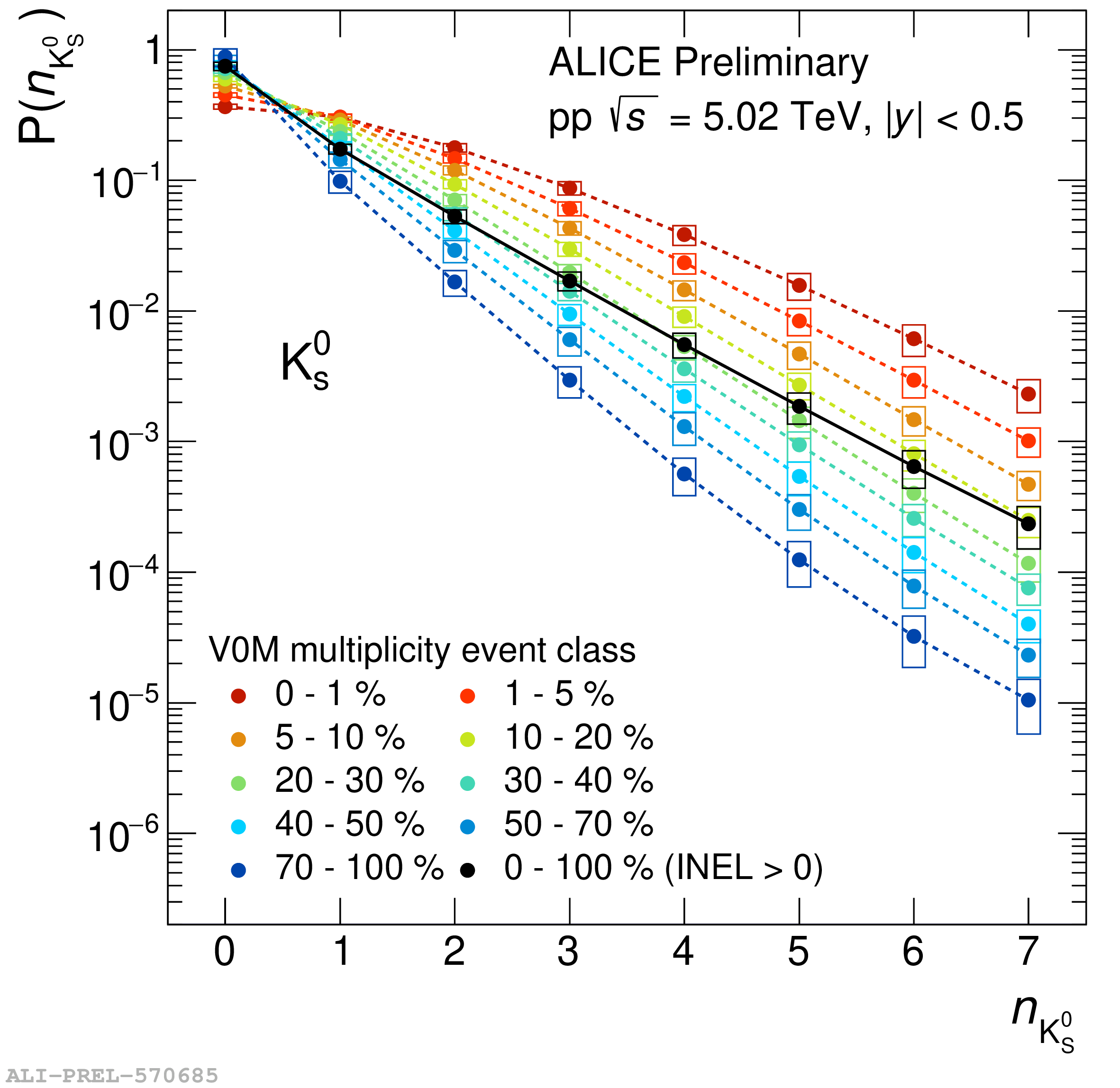} 
    \hspace*{0.2cm}
    \includegraphics[width=0.46 \textwidth, height=0.46\linewidth]{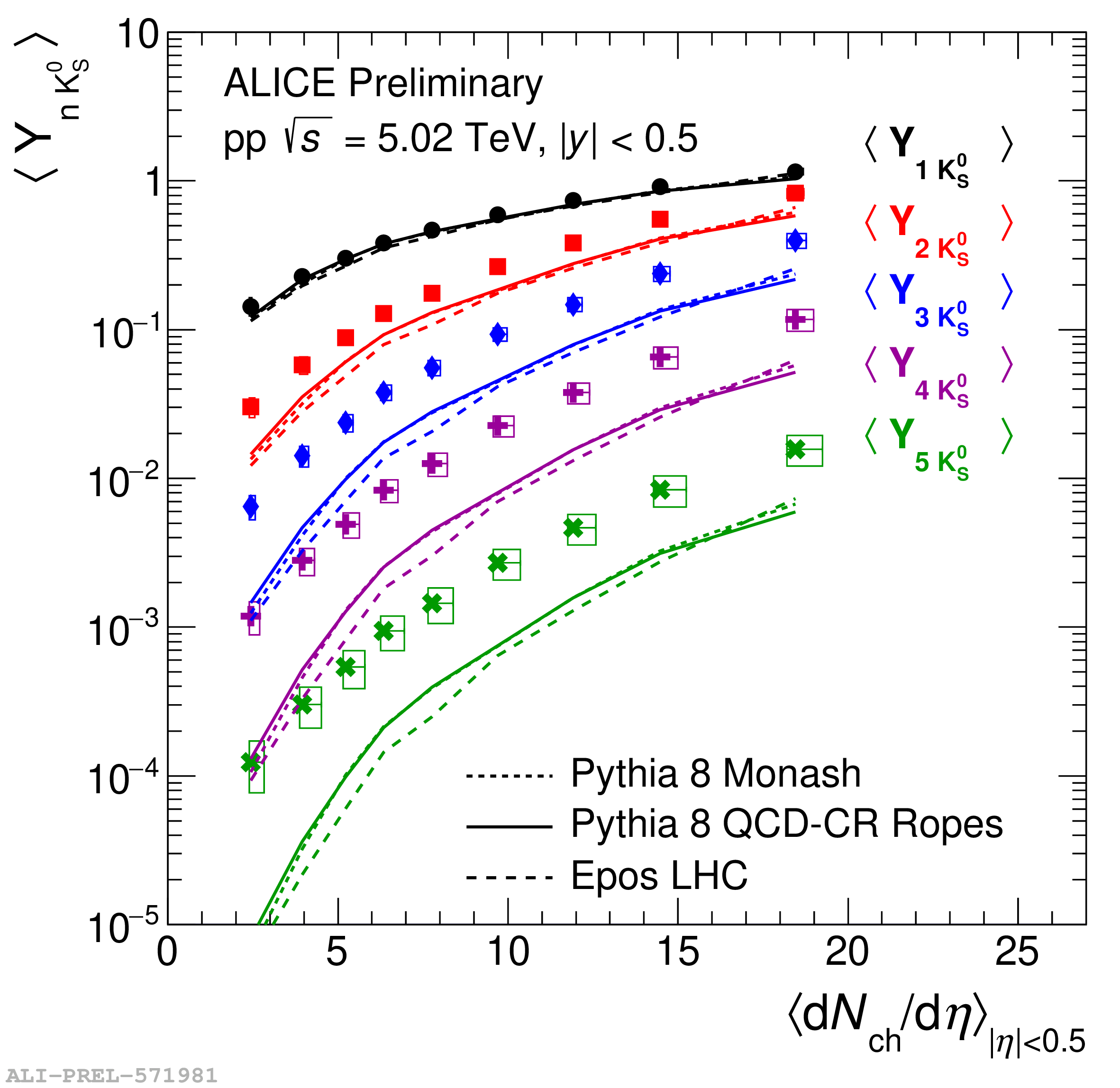}
    \caption{$\textit{Left.}$ ${\rm K}_{\rm S}^{0}$ multiplicity distribution across different event multiplicity classes. The distribution for INEL $>$ 0 events is shown in black, with lines added to guide the eye. $\textit{Right.}$ Multiple strange hadron production yields for ${\rm K}_{\rm S}^{0}$ a function of the charged-particle multiplicity compared with model predictions. Results from PYTHIA 8 Monash, PYTHIA 8 with QCD-CR Ropes and EPOS LHC are represented by dotted, continuous and dashed lines, respectively.}
    \label{fig:Distributions}
\end{figure}

\noindent To deepen the understanding of strangeness production in small systems, ALICE measured event-by-event multiplicity distributions P($\textit{n}_{S}$) for K$^{0}_{S}$, $\Lambda$, $\overline{\Lambda}$, $\Xi^{-}$, $\overline{\Xi}^{+}$, $\Omega$ and $\overline{\Omega}^{+}$ in pp collisions at $\sqrt{s} = 5.02$~\rm{TeV}~\cite{mythesis}. The signal was extracted via a data-driven weighting technique: candidates were assigned a signal probability from invariant mass fits, performed across {\ensuremath{p_{\rm T}}\xspace} and multiplicity bins, using double-sided Crystal Ball~\cite{crystallball} functions and polynomial backgrounds (or a Gaussian for $\Omega$). These probabilities were then combined event-by-event to obtain the P($\textit{n}_{S}$) distributions, in which detector effects were accounted for using a Bayesian unfolding procedure~\cite{unf} applied in one dimension, based on Monte Carlo simulations that realistically model the {\ensuremath{p_{\rm T}}\xspace} distributions and detector performance. For $\Lambda$ and $\overline{\Lambda}$, feed-down contributions from $\Xi^{-}$ ($\overline{\Xi}^{+}$) and $\Xi^{0}$ decays were subtracted. Figure~\ref{fig:Distributions} (left) shows the probability to produce $n{\rm K}_{\rm S}^{0}$ per event for various multiplicity classes, including INEL $>$ 0 category\footnote{INEL $>$ 0 refers to events with at least one reconstructed tracklet in the silicon pixel detector within $|\eta| < 1$, representing approximately 75\% of the inelastic cross section.}. Events with up to 7 ${\rm K}_{\rm S}^{0}$, 5 $\Lambda$, 4 $\Xi$ and 2 $\Omega$ particles per event were observed. Particle–antiparticle yields are consistent across multiplicity, in line with expectations at LHC energies. This novel measurement captures both rare high-strangeness events at low multiplicity and the absence of strange hadrons in high-multiplicity events, providing a unique probe into hadronization dynamics.

\subsection{Multiple strange hadron production yields and yield ratios} 

\noindent From P($\textit{n}{S}$), the average yields $\langle Y{n {\rm -part}} \rangle$ of $n$ strange hadrons per event can be computed via Eq.~\ref{eq:yieldseq}. The right panel of Fig.~\ref{fig:Distributions} shows the average production yields of 1, 2, 3, 4 and 5 ${\rm K}_{\rm S}^{0}$, corresponding to black, red, blue, magenta and green markers, respectively, as a function of the charged-particle multiplicity at midrapidity: a super-linear increase is observed for higher $n$. None of the tested models (Pythia 8 Monash~\cite{Pythia}, Pythia 8 (QCD-CR) Ropes~\cite{PythiaRopes} and Epos LHC~\cite{EPOS}) accurately reproduce the observed trends at high multiplicity.

\begin{equation}
  \big< Y_{nS} \big> = \sum_{i=n}^{\infty} \frac{i!}{n!(i-n)!} \cdot \mathrm{P}(i_{S}).
  \label{eq:yieldseq}
\end{equation}

\begin{figure}
    \begin{center}
    \includegraphics[width=0.99 \textwidth]{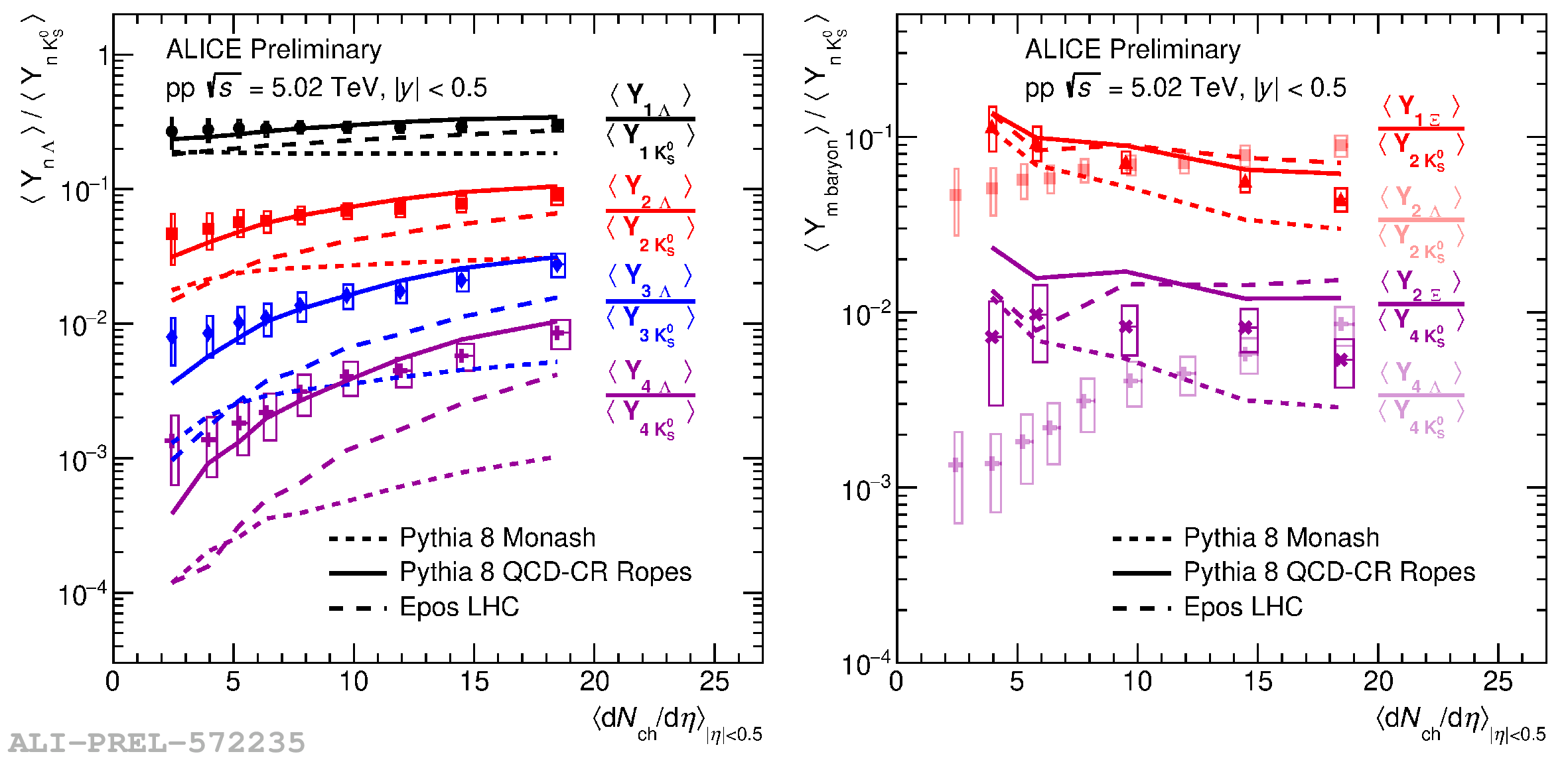}
    \end{center}
    \caption{\textit{Left.} Ratio $\langle Y_{n {\rm \Lambda}} \rangle / \langle Y_{n {{\rm K}_{\rm S}^{0}}} \rangle$ as a function of charged-particle multiplicity. \textit{Right.} $\langle Y_{m {\rm baryons}} \rangle / \langle Y_{n {{\rm K}_{\rm S}^{0}}} \rangle$ as a function of the charged-particle multiplicity. Model predictions from PYTHIA 8 Monash (dotted lines), Pythia 8 QCD-CR Ropes (continuous lines) and Epos LHC (dashed lines) are shown for comparison in both panels. }
    \label{fig:ratiosS0}
\end{figure}

\noindent Additionally, yield ratios with perfect strangeness balance ($\Delta$S = 0) were explored. The $\langle Y_{n {\rm \Lambda}} \rangle / \langle Y_{n {{\rm K}_{\rm S}^{0}}} \rangle$ ratio, reported on the left panel of Fig.~\ref{fig:ratiosS0}, increases with multiplicity, suggesting a growing baryon-to-meson probability. More generally, the ratio of multi-strange baryons to ${\rm K}_{\rm S}^{0}$, illustrated on the right panel of Fig.~\ref{fig:ratiosS0}, decreases with increasing light-quark content, consistent with coalescence expectations: at low multiplicity, excess $s$-quarks favor multi-strange baryons, while at high multiplicity, abundant $u$ and $d$ quarks dominate hadron formation. These findings suggest that, in addition to the difference in the strangeness content between the numerator and the denominator, the evolution with multiplicity is due to the presence of light quarks. Among tested models, Pythia 8 with QCD-CR Ropes~\cite{PythiaRopes} best reproduces the observed trends, suggesting its hadronization mechanism is effective once $s$-quarks are produced, even if the production rate itself remains underestimated.

\section{Conclusions} \label{conclusions}

\noindent Understanding the mechanisms governing strangeness production in small collision systems remains a central objective in high-energy physics. This article presents a coherent overview of recent ALICE results that offer critical insights into this phenomenon from both topological and dynamical perspectives. The analysis of transverse spherocity in conjunction with tight midrapidity multiplicity selections demonstrates that isotropic events favor strangeness production more than jet-like events. These findings suggest that the final-state geometry of the collision plays a significant role, with models incorporating multiple hadronization mechanisms qualitatively reproducing the observed trends. In parallel, the concept of effective energy—derived from ZDC measurements—proves to be a valuable tool for disentangling initial-state effects. The observation that lower forward energy correlates with higher strange hadron yields, even at fixed multiplicity, underscores the importance of the available energy at the partonic level. Furthermore, the measurement of P($\textit{n}_{S}$) distributions represents a valuable probe for testing hadron production models, particularly in events where charged and strange particle yields are unbalanced. Yield ratios involving different strange hadron multiplicities reveal a slight dependence on the number of light quarks and the hadron species. Notably, the increase of $\langle Y_{n {\rm \Lambda}} \rangle / \langle Y_{n {{\rm K}_{\rm S}^{0}}} \rangle$ with multiplicity for $n>1$ suggests that factors beyond simple strangeness conservation—such as quark recombination dynamics—may influence hadron formation. Conversely, decreasing trends in multi-strange baryon ratios hint at the limited role of mass and baryon number in the enhancement mechanism. Altogether, these findings point toward a multifaceted picture of strangeness production in small systems, shaped by both global event properties and local quark dynamics. Future studies involving difference in the strangeness content between the numerator and the denominator greater than three (up to five), combined with improved theoretical modeling, will be instrumental in refining our understanding of QCD in the non-perturbative regime.

\end{document}